\documentclass[prl,aps,amssymb,showpacs,twocolumn]{revtex4-1}
\usepackage[latin1]{inputenc}
\usepackage[T1]{fontenc}
\usepackage{feynmp}
\usepackage[english]{babel}
\usepackage[pdftex]{graphicx}
\usepackage{epstopdf}
\usepackage{amssymb,amsmath}
\usepackage{wasysym}
\DeclareMathOperator{\Tr}{Tr}
\usepackage{tikz}
\usetikzlibrary{trees}
\usetikzlibrary{positioning,arrows}
\usetikzlibrary{decorations.pathmorphing}
\usetikzlibrary{decorations.markings}

\newcommand{\ket}[1]{\left|#1\right\rangle}
\newcommand{\bra}[1]{\left\langle#1\right|}

\newcommand{\up}{\uparrow}
\newcommand{\dw}{\downarrow}

\newcommand{\beq}{\begin{equation}}
\newcommand{\eneq}{\end{equation}}
\input{epsf}

\begin{document}

\tolerance 10000

\newcommand{\vk}{{\bf k}}

%\draft

\title{Triplet FFLO Superconductivity in the doped Kitaev-Heisenberg Honeycomb Model}

\author{Tianhan Liu$^{1,2,4}$}
\author{C\'ecile Repellin$^{3,5}$}
\author{Beno\^{\i}t Dou\c{c}ot$^1$}
\author{Nicolas Regnault$^{3,6}$}
\author{Karyn Le Hur$^2$}
\affiliation{$^1$ Sorbonne Universit\'es, Universit\'e Pierre et Marie Curie, CNRS, LPTHE, UMR 7589 , 4 place Jussieu, 75252 Paris Cedex 05\\ $^2$ Centre de Physique Th\'eorique, \'Ecole polytechnique, CNRS, Universit\'e Paris-Saclay, F-91128 Palaiseau, France\\
$^3$ Laboratoire Pierre Aigrain, Ecole Normale Sup\'erieure-PSL Research
University, CNRS, Universit\'e Pierre et Marie Curie-Sorbonne Universit\'es,
Universit\'e Paris Diderot-Sorbonne Paris Cit\'e, 24 rue Lhomond, 75231
Paris Cedex 05, France\\
$^4$ TCM Group, Cavendish Laboratory, University of Cambridge,
J. J. Thomson Avenue, Cambridge CB3 0HE, United Kingdom\\
$^5$ Max-Planck-Institut f\"ur Physik komplexer Systeme, 01187 Dresden, Germany\\
$^6$ Department of Physics, Princeton university, Princeton NJ 08544, USA.
}

\begin{abstract}

We provide analytical and numerical evidence of a spin-triplet FFLO superconductivity in the itinerant Kitaev-Heisenberg model (anti-ferromagnetic Kitaev coupling and ferromagnetic Heisenberg coupling) on the honeycomb lattice around quarter filling. The strong spin-orbit coupling in our model leads to the emergence of 6 inversion symmetry centers for the Fermi surface at non zero momenta in the first Brillouin zone. We show how the Cooper pairs condense into these non-trivial momenta, causing the spatial modulation of the superconducting order parameter. Applying a Ginzburg-Landau expansion analysis, we find that the superconductivity has  three separated degenerate ground states  with three different spin-triplet pairings. This picture is also supported by exact diagonalizations on finite clusters.
\end{abstract}

\date{\today}

\maketitle

\emph{Introduction}- Mott insulator and high-$T_c$ superconductor are closely related since the latter can be obtained from doping the half-filled Mott insulator \cite{Anderson, F.C.Zhang, Sigrist, Anderson_Lee,Karyn_Maurice}. One key element in superconductivity is the emergence of off-diagonal long-range order which results
in the Bardeen-Cooper-Schrieffer ground state where Cooper pairs have a zero net momentum. The $\eta$ pairing, proposed by C. N. Yang \cite{Yang}, binds electrons with momenta $\mathbf{k}$ and $\pi-\mathbf{k}$, and therefore involves a superconductivity with non-zero Cooper pair momentum. This superconductivity is referred to as the Fulde-Ferrell-Larkin-Ovchinnikov (FFLO) superconductivity \cite{Fulde_Ferrell, Larkin, Larkin1}. The FFLO superconductivity, which supports a spatial modulation for the electron pairing due to the non-trivial Cooper pair momentum, was first proposed in the '60s in a system with significant Zeeman interaction, which shifts the Fermi surfaces for the up and down spins. Experimental realizations of FFLO superconductivity have been proposed, for example, in heavy-fermions \cite{Sarrao}, ultra-cold atom systems \cite{Liao, MartinWolfgang, Zi_Cai, Mora, Chevy_Mora, Koponen}, BEC analogues \cite{Ivana} and in magnetic analogue materials \cite{Green}. However, this exotic phase of matter has not yet been conclusively observed, especially in real materials.

Lately, the studies of ``iridates'', a family of materials with significant spin-orbit coupling, have aroused great interests \cite{KrempaKim, Shitade, Rau} partly because of the emergence of topological Mott physics \cite{pesin} and its connection to the Kitaev anyon model \cite{Kitaev, Halasz}. It has been shown both theoretically and experimentally that the existence of zigzag-magnetic order results from a Kitaev-Heisenberg magnetic coupling in the two-dimensional sodium iridate family \cite{Chaloupka, Jackeli_Khaliullin, Y.Singh, Y.Singh1, Reuther}.  An additional symmetric-off diagonal exchange term can also be added in the analysis \cite{Chun}. Doping these spin-orbit Mott insulators has been addressed theoretically \cite{Rosenow,Ashvin} and has started to attract some experimental attention \cite{Cao}. Here, we address superconductivity in the presence of a large Hubbard interaction and adopt a localized magnetism point of view where the Kitaev-Heisenberg spin Hamiltonian originates from super-exchange processes \cite{Hassan_senechal}. Such Kitaev-Heisenberg physics can also be realized in cold atom systems \cite{Duan}. Using both analytical and numerical methods, we provide convincing evidences of a spin-triplet FFLO superconductor thanks to the spin-orbit coupling close to quarter-filling without breaking the time-reversal symmetry. Superconductivity around quarter-filling has also attracted some attention in the context of graphene \cite{Nandkishore, Annica}.

\begin{figure}[t]
\includegraphics[width=0.445\linewidth]{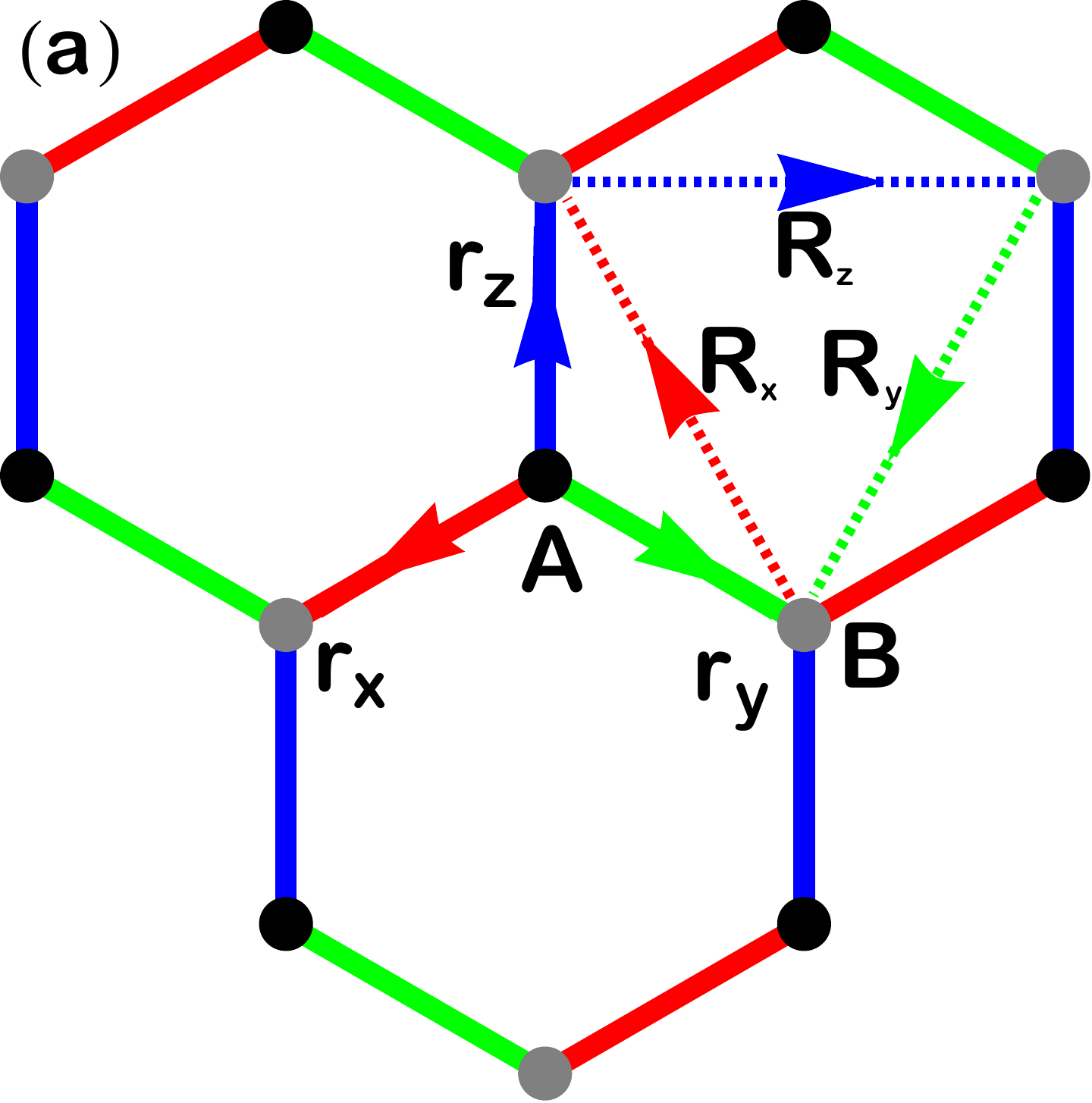}
\includegraphics[width=0.445\linewidth]{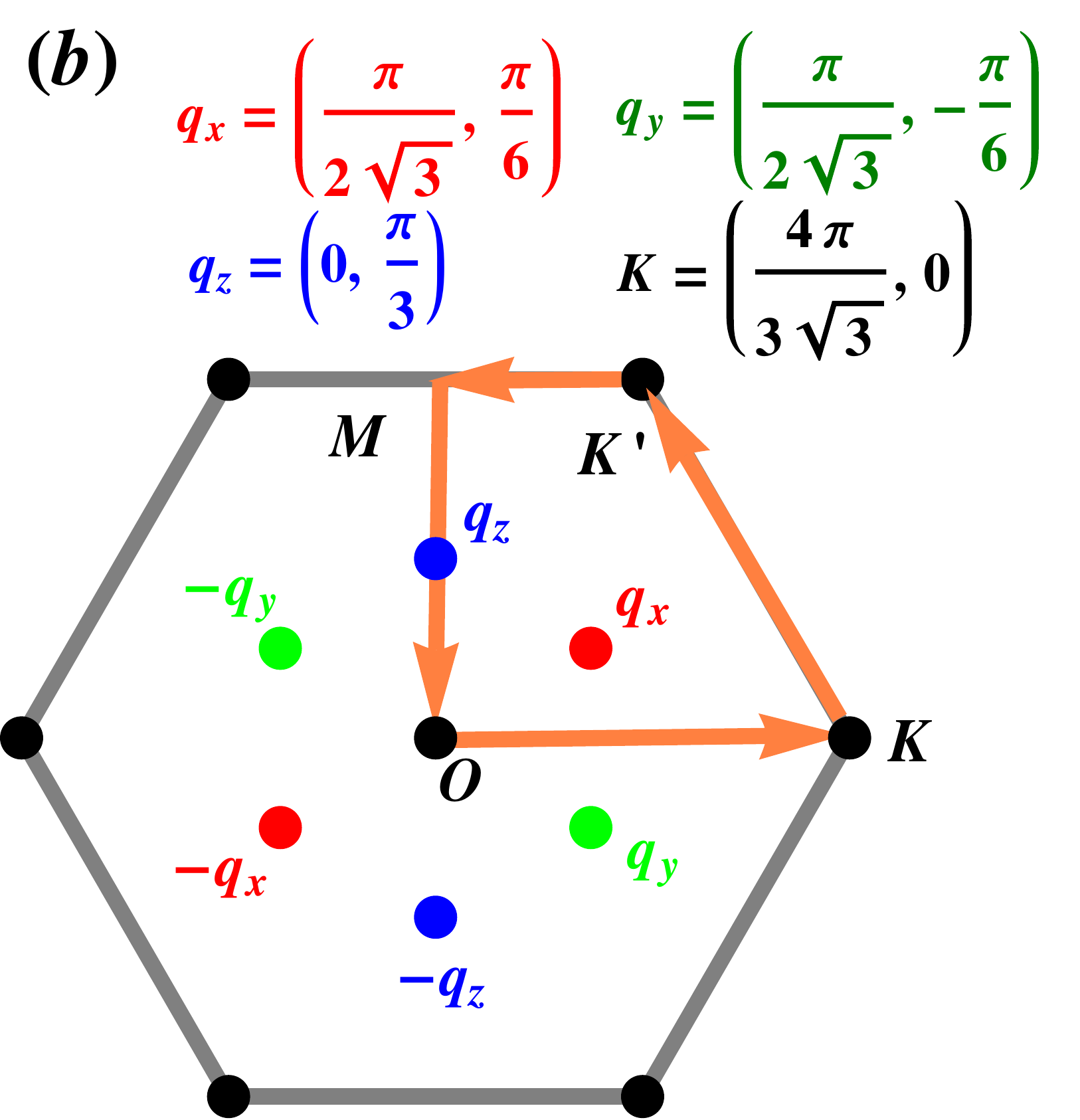}
\includegraphics[width=0.48\linewidth]{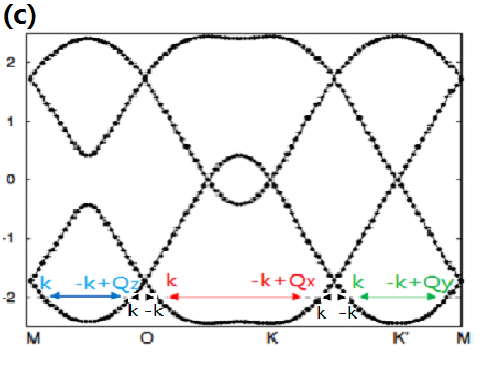}
\includegraphics[width=0.5\linewidth]{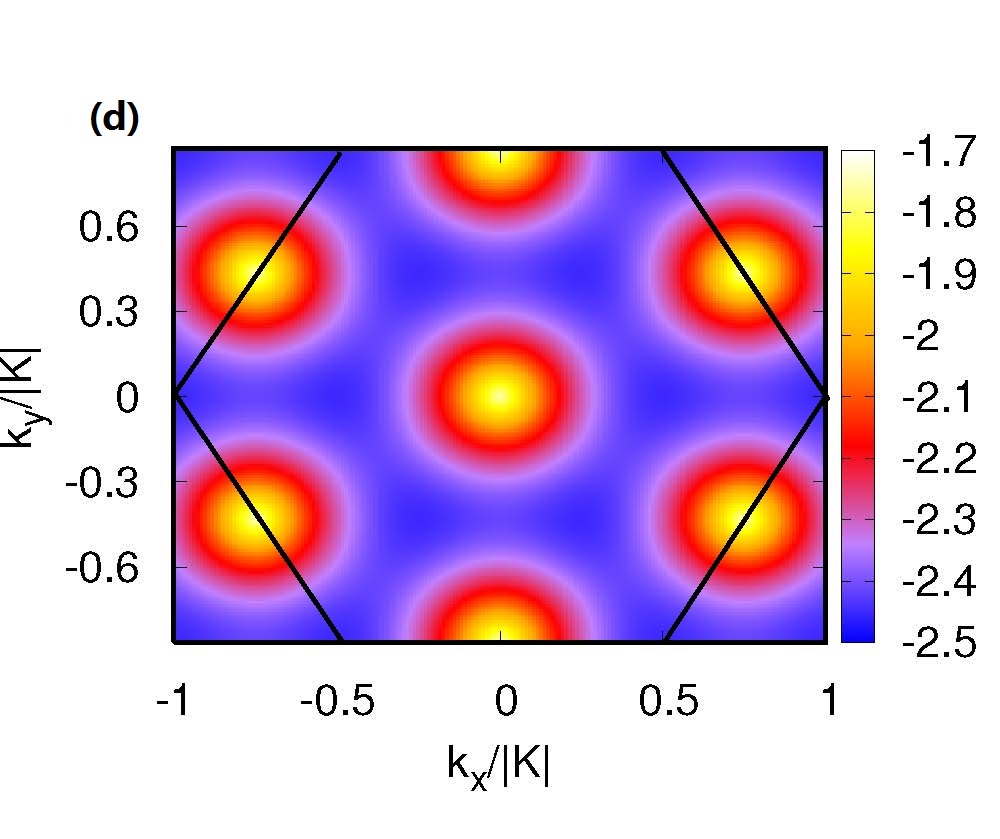}
\caption{(a). The Kitaev-Heisenberg model on the honeycomb lattice: in Eq.~\ref{Hamiltonian} $\alpha$ denotes respectively x on the red links, y on the green links and z on the blue links each of them corresponding to $\mathbf{r}_x=(-\frac{\sqrt{3}}{2}, -\frac{1}{2}); \mathbf{r}_y=(\frac{\sqrt{3}}{2}, -\frac{1}{2}); \mathbf{r}_z=(0, 1);$ the lattice vectors are $\mathbf{R}_x=(-\frac{\sqrt{3}}{2},\frac{3}{2}), \mathbf{R}_y=(-\frac{\sqrt{3}}{2}, -\frac{3}{2}), \mathbf{R}_z=(\sqrt{3}, 0)$. We have taken the lattice spacing to be $1$. (b). The first Brillouin zone, in which, apart from the center of the FBZ, there are six additional centers of inversion symmetry for the Fermi surface of the tight binding part. (c). The band  structure of the spin-orbit model ($t=0, t'=1$). M, O, K, K' are denoted in (b). When the chemical potential is fixed, electrons on the Fermi surface form triplet Cooper pairs with non-trivial momentum $\mathbf{Q}_x$, $\mathbf{Q}_y$ and $\mathbf{Q}_z$. $\mathbf{Q}_{\alpha}=2\mathbf{q}_{\alpha}$. (d) the Fermi surface of the lowest band ($t=0, t'=1$). }\label{psi6}
\end{figure}

Before showing detailed derivations, we summarize the main points. The Kitaev-Heisenberg coupling entails spin-triplet pairing that engenders spinor-condensates \cite{leggett, Zibold, Corre} in momentum space. One important ingredient here is the appearance of 6 inversion symmetry centers for the Fermi surface at non zero momenta in the first Brillouin zone. This will allow the Cooper pairs with triplet pairing to condense at non-trivial momenta. In Fig.~\ref{psi6}, we show the band structure of the spin-orbit coupling model and the symmetry centers of the Fermi surface. Electron pairs around these symmetry centers with non-trivial momenta $\mathbf{q}_{\alpha}$ form spin-triplet pairs with Cooper pair momenta $\mathbf{Q}_{\alpha}=2\mathbf{q}_{\alpha}$. We shall study the superconductivity by calculating the Cooper pairs' response in the Ginzburg-Landau theory for both spin-triplet and spin-singlet pairing. We provide compelling evidence of a triplet FFLO superconductor through  a Ginzburg-Landau expansion and an exact diagonalization analysis.

\emph{Model Hamiltonian}- For the doped Kitaev-Heisenberg model, we consider the following Hamiltonian on the honeycomb lattice: 
\begin{eqnarray}\label{Hamiltonian}
 \begin{split}
 H=&H_0+H_J\\
 H_0=& -\sum_{\left<i,j\right>}P_i[tc_{i\sigma}^{\dagger}d_{j\sigma}+t'c_{i\sigma}^{\dagger}d_{j\sigma'}\tau_{\sigma\sigma'}^{\alpha}+h.c.]P_j\\
 H_J =& J_1\sum_{\left<i,j\right>}\mathbf{S}_i\cdot\mathbf{S}_j+J_2\sum_{\left< i,j\right>}[S_i^{\alpha}S_j^{\alpha}-S_i^{\beta}S_j^{\beta}-S_i^{\gamma}S_j^{\gamma}],
 \end{split}
\end{eqnarray}
here $i$ and $j$ refer to the site index, $c_{i\sigma}$ and $d_{j\sigma}$ to electron operators on the lattices A and B in Fig.~\ref{psi6}a. $\sigma$ and $\sigma'$ are the spins of the electrons and $\tau$ the Pauli matrix with $\alpha=x,y,z$ respectively for red, green and blue links ($\mathbf{r}_i-\mathbf{r}_j=\mathbf{r}_{\alpha}$) and $\beta, \gamma$ take other components than $\alpha$ (See Fig.~\ref{psi6}a).  We note the Gutzwiller projectors as $P_i=(1-\sum_{\sigma}c_{i\sigma}^{\dagger}c_{i\sigma})$ or $P_j=(1-\sum_{\sigma}d_{j\sigma}^{\dagger}d_{j\sigma})$ according to the sub-lattice \cite{Edegger, Rice_Ueda, Rice_Ueda2}. The filling factor $n$ and the doping level $\delta$ are connected by the relation: $n=\frac{1}{2}-\delta$. In contrast to previous analyses \cite{Ashvin, Rosenow}, we include a spin-orbit term of the (doped) model \cite{Hassan_senechal}, such that the anti-ferromagnetic Kitaev and ferromagnetic Heisenberg couplings  at half-filling are microscopically obtained from second-order super-exchange processes: $J_1=\frac{4t^2}{U}, J_2=\frac{4t'^2}{U}$ with U the Hubbard interaction. Setting $J=J_1-J_2$ and $K=J_2$, we recover the model used in Ref.~\cite{Chaloupka} describing the half-filled system.    One shall assume that $t'$ is real to avoid an induced Dzyaloshinskii-Moriya interaction. However, an imaginary $t'$ does not change the physics in the limit of $t=0$. With a purely imaginary $t'$, the time-reversal symmetry (TRS) is restored and we will show the presence of FFLO superconductivity with TRS in this limit.

\begin{figure}[t]
\includegraphics[width=0.325\linewidth]{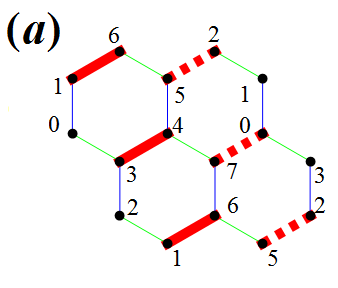}
\includegraphics[width=0.325\linewidth]{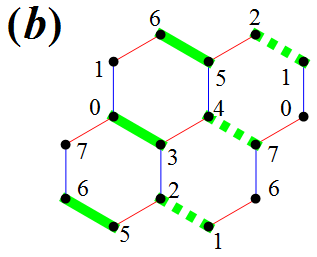}
\includegraphics[width=0.325\linewidth]{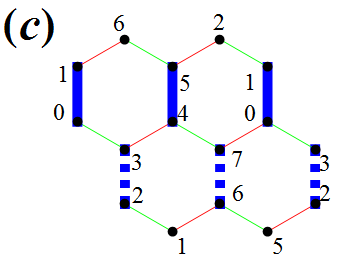}
\caption{The graphical representation of the 3 times degenerate ground state wave function of the FFLO superconductivity around quarter-filling. The bold line signifies a spin-triplet pairing on the link $\Delta_{ij}^{\alpha}$ with the spin-triplet type $\alpha$ (Eq.~\ref{order_parameter}) in correspondence with the type of the link ((a) x red (b) y green and (c) z blue). The dashed line represents the same pairing but with a $\pi$ phase (opposite sign in the wave function).  Here, we only show the nearest-neighbor electron pairing. Long range electron pairing exists and depends on the correlation length of the superconductor \cite{supplement}.}\label{lattice}
\end{figure}

\emph{Band structure around quarter-filling}- Around quarter-filling, which is sufficiently away from half-filling, one can assume that the effect of the Gutzwiller weights on the values of $t'$ is weak and neglect the renormalization of $t'$. We can then diagonalize $H_0$:
\begin{eqnarray}\label{band}
 \begin{split}
 H_0=&\sum_k\Psi_k^{\dagger}\mathcal{H}_{0}(k)\Psi_k, \quad \Psi_k^{\dagger}=(c_{k\up}^{\dagger}, c_{k\dw}^{\dagger}, d_{k\up}^{\dagger}, d_{k\dw}^{\dagger})
\\
 \mathcal{H}_0(\mathbf{k})&=\left(\begin{array}{cc} 0 & M^{\dagger}(\mathbf{k}) \\ M(\mathbf{k}) & 0 \end{array}\right)\\
  M(\mathbf{k})=&t g(\mathbf{k})\tau^{0}+\sum_{\alpha=x, y, z}t'g_{\alpha}(\mathbf{k})\tau^{\alpha}\\
h_{\alpha}(\mathbf{k})=&2t'^2\sin\mathbf{k}\cdot\mathbf{R}_{\alpha}\\
&+2tt'[1+\cos\mathbf{k}\cdot(\mathbf{r}_{\alpha}-\mathbf{r}_{\beta})+\cos\mathbf{k}\cdot(\mathbf{r}_{\alpha}-\mathbf{r}_{\gamma})]
\end{split}
\end{eqnarray}
in which $\alpha\neq \beta, \gamma $ and $g(\mathbf{k})=\sum_{\alpha}e^{i\mathbf{k}\cdot\mathbf{r}_{\alpha}}$, and $g_{\alpha}(\mathbf{k})=e^{i\mathbf{k}\cdot\mathbf{r}_{\alpha}}$ ($\alpha=x, y, z$). We see that in the spin-orbit coupling limit ($t=0$) the Fermi surface has six additional inversion symmetry centers, apart from the inversion symmetry center $O$ with trivial momentum $\mathbf{Q}_0=0$, in the first Brillouin zone (FBZ) $\mathbf{k}\leftrightarrow 2\mathbf{q}_{\alpha}-\mathbf{k}$  ($\alpha=x, y, z$) as indicated in Fig.~\ref{psi6}b. This derives from the Sine function remaining invariant under the change of $\mathbf{k}\cdot\mathbf{R}_{\alpha}\leftrightarrow \pi -\mathbf{k}\cdot\mathbf{R}_{\alpha}$. In Fig.~\ref{psi6}c, we show the band structure  at the spin-orbit coupling limit $t=0, t'\neq0$: the four bands have a conic structure for the Fermi surface at half and quarter filling.

{ \emph{Superconducting Instability}}- The doped itinerant Kitaev-Heisenberg model in the spin-orbit limit ($t=0$) has $7$ symmetry centers  around quarter-filling with momenta: $\pm\mathbf{q}_{\alpha}$ ($\alpha=x, y, z$) and $\mathbf{q}_0=0$. There are 4 kinds of Cooper pairs around these symmetry centers \cite{Center symmetry, supplement}:
\begin{equation}\label{order_parameter_pairing}
\hat{\Delta}^{\dagger}_{\alpha\mathbf{Q}_{\alpha}}(\mathbf{k})=i\tau_{\sigma\sigma"}^y\tau_{\sigma"\sigma'}^{\alpha}c_{\mathbf{k}\sigma}^{\dagger}d_{-\mathbf{k}+\mathbf{Q}_{\alpha}\sigma'}^{\dagger} \quad (\alpha=0, x, y, z)
\end{equation}
In the direct space, the three types of spin-triplet pairing and the spin-singlet pairing in competition are:
\begin{eqnarray}\label{order_parameter}
\begin{split}
&\hat{\Delta}_{ij}^x =c_{i\uparrow}d_{j\uparrow}-c_{i\downarrow}d_{j\downarrow};  \hat{\Delta}_{ij}^y=i( c_{i\uparrow}d_{j\uparrow}+c_{i\downarrow}d_{j\downarrow} );\\ &\hat{\Delta}_{ij}^z=c_{i\uparrow}d_{j\downarrow}+c_{i\downarrow}d_{j\uparrow}; \hat{\Delta}_{ij}^0=c_{i\uparrow}d_{j\downarrow}-c_{i\downarrow}d_{j\uparrow}.
\end{split}
\end{eqnarray}
The Kitaev-Heisenberg coupling involves the density channel $\hat{\chi}_{\alpha}=c_{i\sigma}^{\dagger}d_{j\sigma'}\tau_{\sigma\sigma'}^{\alpha}+h.c.$ besides the superconductivity pairing. We have checked that around quarter-filling the density channel renormalizes the spin-orbit coupling term $t'$ and such renormalization is negligible \cite{supplement}. Then we can decompose the Kitaev-Heisenberg coupling at the mean-field level as:
\begin{eqnarray}\label{mean_field}
\begin{split}
&J_2\sum_{\left< i,j\right>}[S_i^{\alpha}S_j^{\alpha}-S_i^{\beta}S_j^{\beta}-S_i^{\gamma}S_j^{\gamma}]\\
=&\frac{3J_2N_s}{4}\sum_{\alpha, \mathbf{Q}}|\Delta_{\alpha\mathbf{Q}}|^2-J_2\sum_{\alpha, \mathbf{k}, \mathbf{Q}}[g_{\alpha}(\mathbf{k})\Delta_{\alpha\mathbf{Q}}\hat{\Delta}_{\alpha\mathbf{Q}}^{\dagger}(\mathbf{k})\\
&-g_{\alpha}(\mathbf{k})\Delta_{0\mathbf{Q}}\hat{\Delta}_{0\mathbf{Q}}^{\dagger}(\mathbf{k})+h.c.],
\end{split}
\end{eqnarray}
in which $\Delta_{\alpha\mathbf{Q}}=\frac{1}{N_s}\sum_{\left<i, j\right>}e^{i\mathbf{Q}\cdot\mathbf{r}_j}\left<\hat{\Delta}_{ij}^{\alpha}\right>$ is the Fourier transform of the order parameter $\left<\hat{\Delta}_{ij}^{\alpha}\right>$ in Eq.~\ref{order_parameter}  with spatial phase modulation  $e^{i\mathbf{Q}\cdot\mathbf{r}_j}$. $N_s$ denotes here the number of unit cells.

\begin{figure}[tp]
\includegraphics[width=0.999\linewidth]{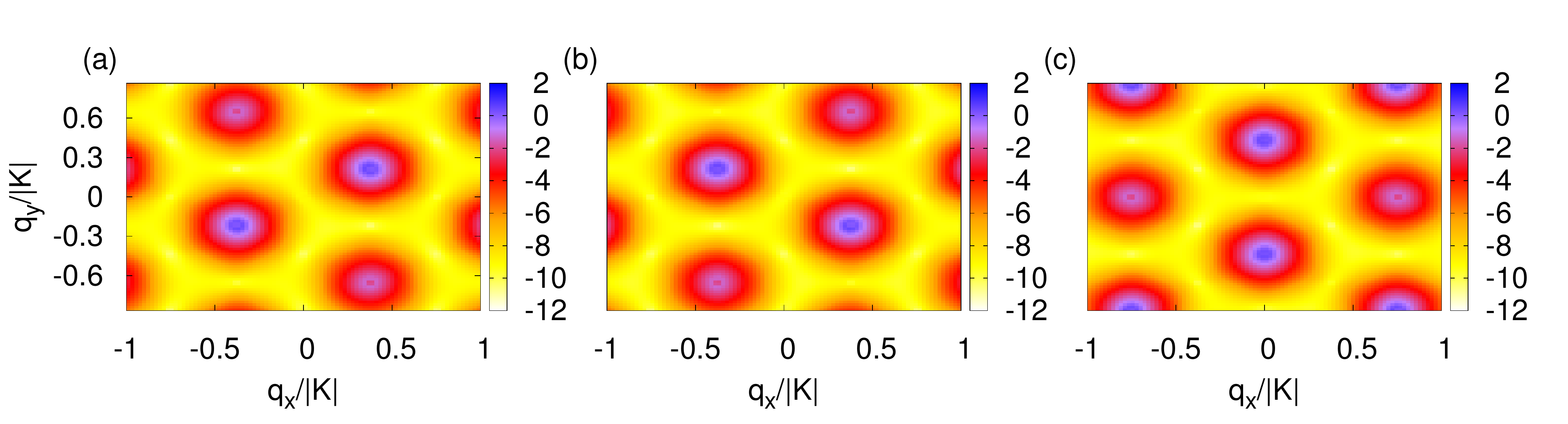}
\caption{The inverse of the vertex function $\Gamma^{-1}_{\alpha}(\mathbf{Q}, T)$ at quarter-filling $\alpha=x, y, z$ as a function of $\mathbf{q}=\frac{\mathbf{Q}}{2}\in$ FBZ (first Brillouin zone) at quarter-filling for the spin-triplet pairing $\Delta_{x\mathbf{Q}}$ (a), $\Delta_{y\mathbf{Q}}$ (b) and $\Delta_{z\mathbf{Q}}$ (c) at temperature $k_BT=0.01t'$ and $t'=1$. }\label{susc_fbz1}
\end{figure}

We constitute the Nambu spinor for the four Cooper pairs  $\Phi_{\mathbf{k}\mathbf{Q}}=(\Psi_{\mathbf{k}}, \Psi_{\mathbf{Q}-\mathbf{k}}^{\dagger})$ ($\Psi_{\mathbf{k}}$ is defined in Eq.~\ref{band}) and write down their Gor'kov-Green function $G_{\alpha}^{-1}(\omega, \mathbf{k}, \mathbf{Q})$ ($\alpha=0, x, y, z$, $\mathbf{Q}/2 \in$  FBZ). We then pursue the Landau expansion \cite{supplement}. In the spin-orbit coupling limit ($t=0$, $t'\neq0$),  we have the second order  Landau expansion (here we fix $U=6$ following Ref.~\cite{Rosenow}):
\begin{equation}\label{Landau}
F_{BCS}\approx-\sum_{\alpha, \beta=0, x, y, z}\sum_{\mathbf{Q}}N_s\Gamma^{-1}_{\alpha\beta}(\mathbf{Q}, T)\Delta_{\alpha\mathbf{Q}}\Delta_{\beta\mathbf{Q}}^*
\end{equation}
in which $F_{BCS}$ is  the free energy and to the lowest (second) order is proportional to the inverse of the Cooper pair vertex function $\Gamma^{-1}_{\alpha\beta}(\mathbf{Q}, T)$ \cite{supplement}. When $\alpha\neq\beta$, we have checked that $\Gamma^{-1}_{\alpha\beta}(\mathbf{Q}, T)$ is negligible because of frustration in the momentum space; therefore we focus our attention on the diagonal part of the inverse of the Cooper pair vertex function that we denote as $\Gamma^{-1}_{\alpha}(\mathbf{Q}, T)\equiv\Gamma^{-1}_{\alpha\alpha}(\mathbf{Q}, T)$.  When $\Gamma^{-1}_{\alpha}(\mathbf{Q}, T)>0$, the triplet superconductor pairing $\Delta_{\alpha\mathbf{Q}}$ is stable \cite{AltlandSimons}. In Fig.~\ref{susc_fbz1}, we show   $\Gamma^{-1}_{\alpha}(\mathbf{Q}, T)$ as a function of $\mathbf{q}=\mathbf{Q}/2\in$ FBZ at temperature $k_BT=0.01t'$, in which we remark the condensation of spin-triplet Cooper pairs $\Delta_{\alpha\mathbf{Q}}$ into the peaks at wave vector $\mathbf{q}_{\alpha}=\frac{\mathbf{Q}_{\alpha}}{2}$. We have three spin-triplet condensates at different momenta as shown in Fig.~\ref{lattice} a, b and c.

\begin{figure}[htb]
\begin{center}
\includegraphics[width=0.48\linewidth]{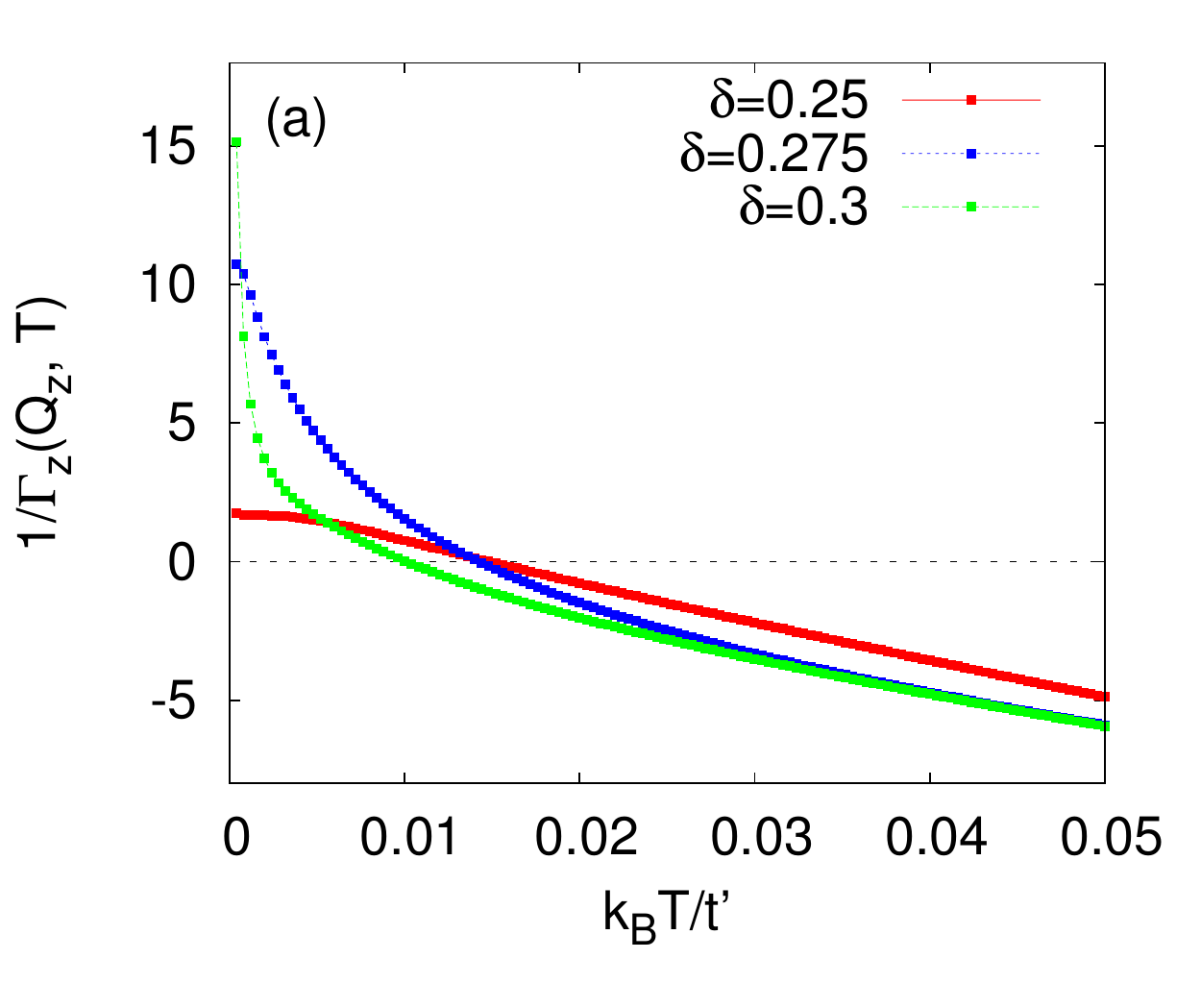}
\includegraphics[width=0.48\linewidth]{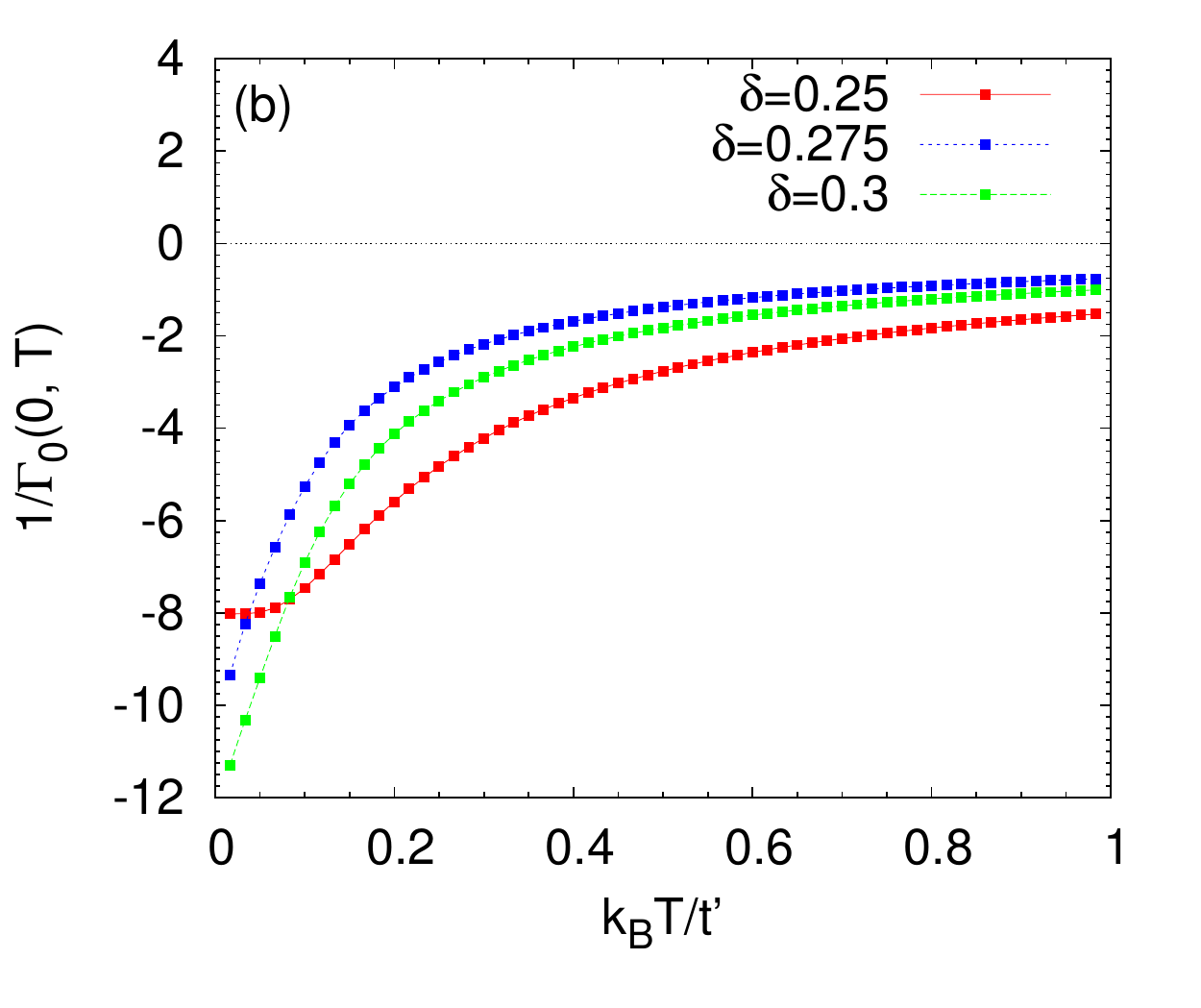}
\caption{ In the limit $t=J_1=0$: (a) The peak of Cooper pair vertex function $\Gamma^{-1}_{\alpha}(\mathbf{Q}_{\alpha}, T)$ as a function of temperature at different doping level ($\delta=0.25$ is the quarter-filling). (b) The vertex function of singlet Cooper pair $\Gamma^{-1}_{0}(0, T)$ as a function of temperature at different doping level. }\label{susc_dop}
\end{center}
\end{figure}

We also study the peak of the static Cooper pair response $\Gamma^{-1}_{\alpha}(\mathbf{Q}_{\alpha}, T)$ as a function of temperature at different doping levels $\delta$: the peak remains finite at quarter-filling, while it has logarithmic divergence at zero temperature when the doping diverts from quarter-filling (Fig.~\ref{susc_dop}a). Here, $\Gamma^{-1}_{\alpha}(\mathbf{Q}_{\alpha}, T)$ is proportional to the density of states at the Fermi level,  which vanishes linearly as $\delta\rightarrow1/4$, which means that at quarter-filling superconductivity disappears and we have a free electron system, assuming $J_2$ is not too large compared to $t'$. At low temperature, the peak of the condensate profile $\Gamma^{-1}_{x}(\mathbf{Q}_{x}, T)=\Gamma^{-1}_{y}(\mathbf{Q}_{y}, T)=\Gamma^{-1}_{z}(\mathbf{Q}_{z}, T)$ stays positive while the peak of the spin-singlet condensate profile $\Gamma^{-1}_{0}(0, T)$ remains negative at all temperature (Fig.~\ref{susc_dop}b). This indicates that in the spin-orbit coupling limit, the doped itinerant Kitaev-Heisenberg model hosts only the three spin-triplet ground states.

The three spin-triplet condensates may interact with each other and we have calculated the box diagram in Fig.~\ref{box_diagram} to study this effect by extending the Landau expansion to the fourth order. In Fig.~\ref{box_diagram}, we note 
\begin{eqnarray}\label{boson}
\begin{split}
&b_{xq}^{\dagger}=\frac{1}{N_s}\sum_{\mathbf{k}}(c_{\mathbf{k}\uparrow}^{\dagger}d_{-\mathbf{k}+\mathbf{q}\uparrow}^{\dagger}-c_{\mathbf{k}\downarrow}^{\dagger}d_{-\mathbf{k}+\mathbf{q}\downarrow}^{\dagger})\\
&b_{yq}^{\dagger}=-i\frac{1}{N_s}\sum_{\mathbf{k}}( c_{\mathbf{k}\uparrow}^{\dagger}d_{-\mathbf{k}+\mathbf{q}\uparrow}^{\dagger}+c_{\mathbf{k}\downarrow}^{\dagger}d_{-\mathbf{k}+\mathbf{q}\downarrow}^{\dagger} )\\
&b_{zq}^{\dagger}=-\frac{1}{N_s}\sum_{\mathbf{k}}( c_{\mathbf{k}\uparrow}^{\dagger}d_{-\mathbf{k}+\mathbf{q}\downarrow}^{\dagger}+c_{\mathbf{k}\downarrow}^{\dagger}d_{-\mathbf{k}+\mathbf{q}\uparrow}^{\dagger} )
\end{split}
\end{eqnarray}
as the creation operators for the three Cooper pairs. Since the three Cooper pairs condense at different momenta $\mathbf{Q}_{\alpha}$, the box diagram is actually the only one respecting momentum conservation. To the fourth order, we obtain the free energy of the three condensates:
\begin{eqnarray}
F_{BCS}=&&N_s\sum_{\alpha=x, y, z}\{-\Gamma^{-1}_{\alpha}(\mathbf{Q}_{\alpha}, T)|\Delta_{\alpha\mathbf{Q}_{\alpha}}|^2+C_1|\Delta_{\alpha\mathbf{Q}_{\alpha}}|^4\}\nonumber\\
&&+N_sC_2\sum_{\alpha\neq\beta}|\Delta_{\alpha\mathbf{Q}_{\alpha}}|^2|\Delta_{\beta\mathbf{Q}_{\beta}}|^2,
\end{eqnarray}
in which $C_1$ and $C_2$ are positive numbers obtained from the calculation of the box diagram in Fig.~\ref{box_diagram}. We have checked that $C_2>C_1>0$ and thus we deduce that mixing of the three superconducting condensates is not energetically favorable, and there is phase separation among the three types of fermionic pairs. Consequentially, the ground state wave function at zero temperature is three times degenerate (See Fig.~\ref{lattice}): the modulated $\Delta_{ij}^{\alpha}$ (Eq.~\ref{order_parameter}) are represented by bold and dashed lines ((a) red for X,  (b) green for Y and (c) blue for Z).

When $t$ and $J_1$ are small compared to $t'$ and $J_2$, the three FFLO states are still stable when the temperature is low enough  ($\Gamma^{-1}_{x}(\mathbf{Q}_{x}, T)=\Gamma^{-1}_{y}(\mathbf{Q}_{y}, T)=\Gamma^{-1}_{z}(\mathbf{Q}_{z}, T)>0)$ . The FFLO phase remains stable as long as the energy related to the critical temperature is bigger than the gap of the free electron system around quarter-filling opened by the $t$ term i.e. $k_BT_c(\delta)>t$.

\begin{center}
\tikzset{
particle/.style={thick,draw=blue, postaction={decorate},
    decoration={markings,mark=at position .5 with {\arrow[blue]{triangle 45}}}},
gluon/.style={decorate, draw=black,
    decoration={coil,aspect=0}},
photon/.style={decorate, decoration={snake}, draw=red},
fermion/.style={draw=black, postaction={decorate},decoration={markings,mark=at position .55 with {\arrow{>}}}},
  vertex/.style={draw,shape=circle,fill=black,minimum size=3pt,inner sep=0pt}
 }

\begin{figure}
\begin{tikzpicture}[node distance=1cm and 1.5cm]
\coordinate[label=left:$b_{\alpha\mathbf{Q}_{\alpha}}$] (e1);
\coordinate[below right=of e1] (aux1);
\coordinate[right=of aux1] (aux2);
\coordinate[above right=of aux2,label=right:$\bar{b}_{\alpha\mathbf{Q}_{\alpha}}$] (e2);
\coordinate[below=1.25cm of aux1] (aux3);
\coordinate[below=1.25cm of aux2] (aux4);
\coordinate[below left=of aux3,label=left:$\bar{b}_{\beta\mathbf{Q}_{\beta}}$] (e3);
\coordinate[below right=of aux4,label=right:$b_{\beta\mathbf{Q}_{\beta}}$] (e4);

\draw[photon] (e1) -- (aux1);
\draw[particle] (aux1) -- (aux2) node[midway,above=0.1cm] {$\mathbf{p}$};
\draw[particle] (aux1) -- (aux3) node[midway,left=0.1cm] {-$\mathbf{p}+\mathbf{Q}_{\alpha}$};
\draw[particle] (aux4) -- (aux2) node[midway,right=0.1cm] {-$\mathbf{p}+\mathbf{Q}_{\alpha}$};
\draw[photon] (aux2) -- (e2);
\draw[photon] (aux3) -- (e3);
\draw[particle] (aux4) -- (aux3) node[midway,below=0.1cm] {$\mathbf{p}-\mathbf{Q}_{\alpha}+\mathbf{Q}_{\beta}$};
\draw[photon] (e4) -- (aux4);
\end{tikzpicture}
\caption{The box diagram of the $4th$ order Landau expansion describing the interaction between the triplet pairing.}\label{box_diagram}
\end{figure}
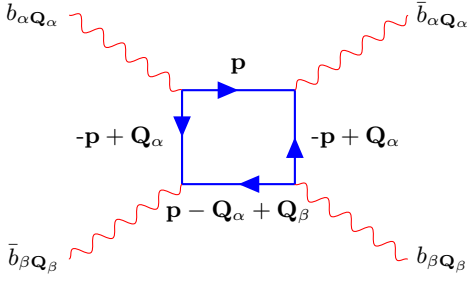
\end{center}

\begin{figure}[th]
\includegraphics[width=0.96\linewidth]{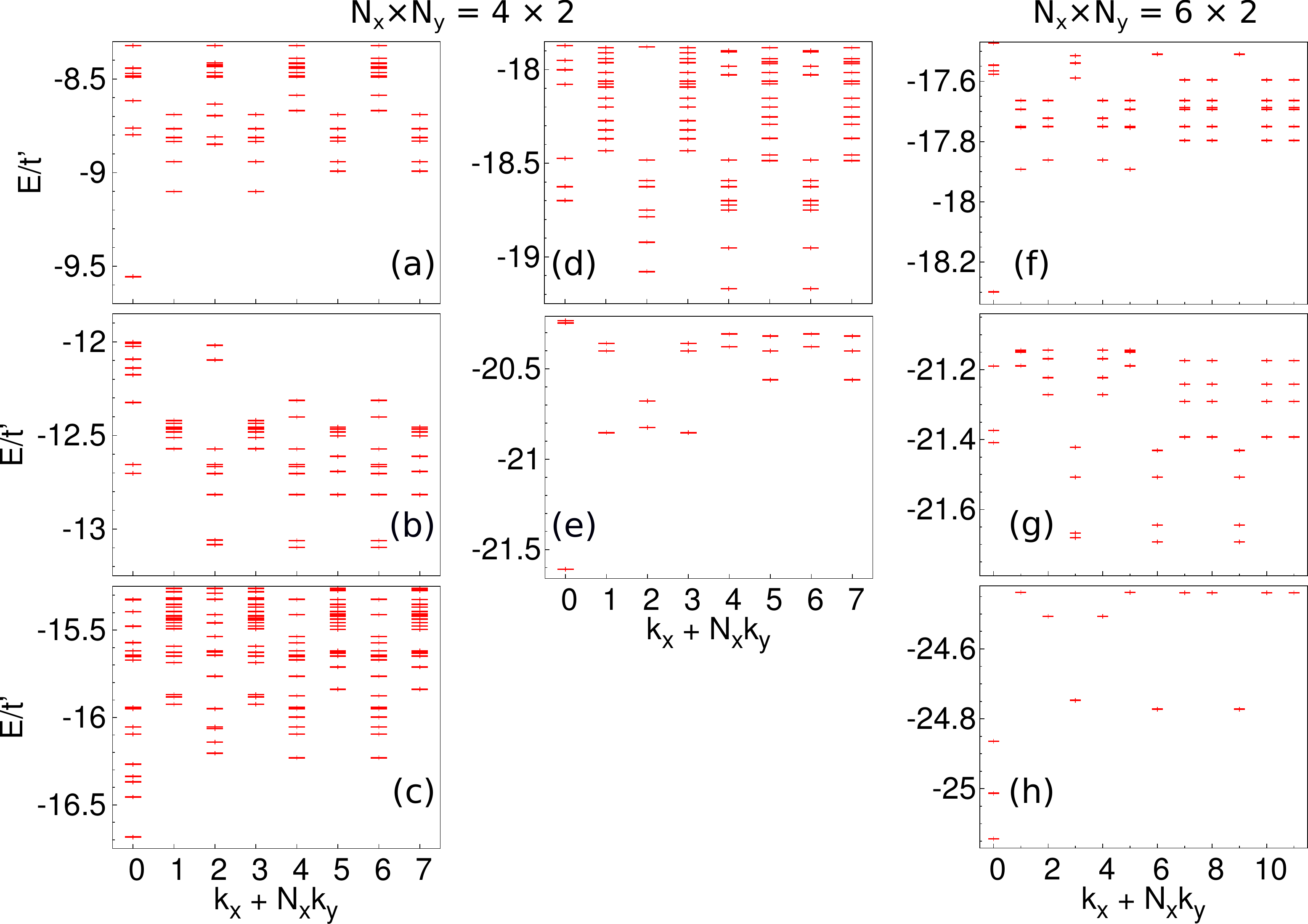}
\caption{Energy spectra as a function of the linearized momentum $k_x + N_x k_y$ of the Hamiltonian in Eq.~\ref{Hamiltonian} with periodic boundary conditions and $t=J_1=0$, $t'=1, J_2=0.667$. The left column and middle column show a system of $N_x\times N_y=4\times 2$ plaquettes with particle numbers (a) $N=4$ (b) $N=6$ (c) $N=8$ or quarter-filling (d) $N=10$ and (e) $N=12$. The right column only provides the (f) $N=8$, (g) $N=10$, (h) $N=12$ spectra on a $6\times 2$ system (the largest Hilbert space dimension involved for (h) is $\simeq 1.7 . 10^{8}$). Note that for this system, only the few first energy levels are shown.}\label{dop}
\end{figure}

\emph{Exact Diagonalization of the Kitaev-Heisenberg model}- We have done an exact diagonalization of the Kitaev-Heisenberg model of Eq.~\ref{Hamiltonian} in the spin-orbit coupling limit $t=0, t'=1$. The exact diagonalization treats the Gutzwiller projectors exactly in Eq.~\ref{Hamiltonian}. We fix the parametrization $J_1=\frac{4t^2}{U}$, $J_2=\frac{4t'^2}{U}$ (here we choose $U=6$ as suggested by Ref.~\cite{Rosenow}). The system has $N_s=N_x\times N_y$ plaquettes with periodic boundary conditions in both directions, and is filled with $N$ electrons on the $2N_x\times N_y$ sites.
$N_x$ and $N_y$ are both even numbers in order to avoid frustration of the FFLO condensates. Due to computational constraints, we reduce our study to three system sizes: $N_y = 2$, $N_x = 2,\ 4,\ 6$. For an odd number of Cooper pairs (doped system), the lowest energy eigenstates appear in momentum sectors $\mathbf{k}_x=(N_x/2, N_y/2)$ $\mathbf{k}_y=(N_x/2, 0)$ and $\mathbf{k}_z=(0, N_y/2)$ (in the bases of $\mathbf{k}_1=\frac{1}{N_x}(0, \frac{4\pi}{3})$, $\mathbf{k}_2=\frac{1}{N_y}(\frac{2\pi}{\sqrt{3}}, -\frac{2\pi}{3})$.) as shown in Figs.~\ref{dop}b, d, and g.  The degeneracy for the three spin-triplet states is partially lifted when $N_x \neq N_y$ which breaks the symmetry of a $2\pi/3$ rotation followed by a permutation of spin components. For an even number of Cooper pairs, the ground state appears in momentum sector $\mathbf{k}_{0}= (0, 0)$ as shown in Figs.~\ref{dop}a, c, e, f and h. 
In agreement with the theory, $\mathbf{k}_{\alpha}$ coincides with the three discrete version of the FFLO Cooper pair momenta $\mathbf{q}_{\alpha}$ ($\alpha=x, y, z$)  for an odd number of Cooper pairs while for even number of Cooper pairs $\mathbf{k}_0\equiv 2\mathbf{k}_{\alpha}=2\mathbf{q}_{\alpha}\mod(N_x, N_y)$.  This alternation of ground state momentum sector as a function of particle numbers distinguishes the FFLO superconductivity here from other modulated orders like spin or charge density waves \cite{supplement}. The quasi-degeneracy in Fig.\ref{dop}b is yet to be understood and might just be a finite size effect.

\emph{Conclusion}- We have provided both analytical and numerical evidence of a pure spin-triplet FFLO superconductor in the doped itinerant Kitaev-Heisenberg model in the spin-orbit coupling limit ($t, J_1 \rightarrow0$). When $t'$ is purely imaginary, the time-reversal symmetry (TRS) is restored. The key ingredient of the FFLO superconductivity here is the symmetry centers of the Fermi surface at non-trivial momenta instead of a Zeeman field. The ground state is three times degenerate with respectively the three spin-triplet pairing $\Delta_{ij}^{\alpha}$ in the p-wave state with non-trivial Cooper pair momentum $\mathbf{Q}_{\alpha}=2\mathbf{q}_{\alpha}$ and spatial moduation of $\pi$ phase in the direction of lattice vector $\mathbf{R}_{\alpha}$ for the order parameter.  These results may have relevance for doped iridate honeycomb materials or in ultra-cold atom systems.

\emph{Acknowledgments}- We acknowledge discussions with Sylvain Capponi,  Claudio Castelnovo, Fabrice Gerbier,  Loic Herviou,  Dmitry Kovrizhin, Claudine Lacroix, Philippe Lecheminant, Fr\'ed\'eric Mila,   Christophe Mora,   Catherine P\'epin, Alexandru Petrescu, Didier Poilblanc,  and Julien Vidal. N.R. was supported by the Princeton Global Scholarship. K.L.H. has benefited from discussions at KITP Santa-Barbara and CIFAR meetings in Canada, and was supported in part by the National Science Foundation under Grant No. PHY11-25915.

\newpage

\begin{widetext}

\section{Supplemental material}

In the supplementary material, we provide additional analytical and numerical results that might be relevant to a more specialized audience.

\subsection{Spin-Triplet Pairings}

In the weak coupling limit (free electrons), we can measure the spin of the 4 electron pairs as shown in Fig.~\ref{psi6}a at the Fermi level using the projector $P_{sp}(\mathbf{k})=\frac{1}{2}[1-p\hat{\vec{h}}(\mathbf{k})\cdot\vec{\tau}], \tilde{\vec{h}}(\mathbf{k})=\vec{h}(\mathbf{k})/|\vec{h}(\mathbf{k})|$ as $\left< \mathbf{S}_k\right>=\Tr(P_{sp}(\mathbf{k})\hat{\mathbf{S}}P_{sp}(\mathbf{k}))$ ($p=1$ for the first and fourth band $p=-1$ for the second and third band). The functions $h_{\alpha}(\mathbf{k})$ are given in Eq.~\ref{band}. In the limit of $t\rightarrow 0$ if we measure the spin of the two electron pairs at the Fermi level connected by $\mathbf{Q}_x$ that we denote as $S^x_{\mathbf{k}}$ and $S^x_{\mathbf{Q}_x-\mathbf{k}}$,  we have:
\begin{equation}\label{spin_pairing_x}
\left< S^{x}_{\mathbf{k}}\right>=-\left< S^{x}_{\mathbf{Q}_{x}-\mathbf{k}}\right>\quad \left< S^{y}_{\mathbf{k}}\right>=\left< S^{y}_{\mathbf{Q}_{x}-\mathbf{k}}\right>\quad \left< S^{z}_{\mathbf{k}}\right>=\left< S^{z}_{\mathbf{Q}_{x}-\mathbf{k}}\right>
\end{equation}

Similarly for the spin expectation values of the electron pairs connected by $\mathbf{Q}_y$ and $\mathbf{Q}_z$ have the following relation:
\begin{eqnarray}\label{spin_pairing_yz}
\begin{split}
&\left< S^{y}_{\mathbf{k}}\right>=-\left< S^{y}_{\mathbf{Q}_{y}-\mathbf{k}}\right>\quad \left< S^{z}_{\mathbf{k}}\right>=\left< S^{z}_{\mathbf{Q}_{y}-\mathbf{k}}\right>\quad \left< S^{x}_{\mathbf{k}}\right>=\left< S^{x}_{\mathbf{Q}_{y}-\mathbf{k}}\right> \\
&\left< S^{z}_{\mathbf{k}}\right>=-\left< S^{z}_{\mathbf{Q}_{z}-\mathbf{k}}\right>\quad \left< S^{x}_{\mathbf{k}}\right>=\left< S^{x}_{\mathbf{Q}_{z}-\mathbf{k}}\right>\quad \left< S^{y}_{\mathbf{k}}\right>=\left< S^{y}_{\mathbf{Q}_{z}-\mathbf{k}}\right>, 
\end{split}
\end{eqnarray}

In particular, for the electron pairs with zero Cooper pair momentum, we have:
\begin{equation}
\left< \mathbf{S}_{\mathbf{k}}\right>=-\left< \mathbf{S}_{-\mathbf{k}}\right>.
\end{equation}

We can see obviously from Eq.~\ref{spin_pairing_x} and Eq.~\ref{spin_pairing_yz} that electron pairs around symmetry center $\mathbf{q}_{\alpha}$ minimize the coupling in the form of Kitaev-Heisenberg $J\left<(S_{\mathbf{k}}^{\alpha}S_{\mathbf{Q}_{\alpha}-\mathbf{k}}^{\alpha}-S_{\mathbf{k}}^{\beta}S_{\mathbf{Q}_{\alpha}-\mathbf{k}}^{\beta}-S_{\mathbf{k}}^{\gamma}S_{\mathbf{Q}_{\alpha}-\mathbf{k}}^{\gamma})\right>=-J\left<\mathbf{S}_k^2\right>$ and the electron pairs around the inversion center $O$ minimize the coupling in the form of Heisenberg $J\left<\mathbf{S}_{\mathbf{k}}\cdot\mathbf{S}_{-\mathbf{k}}\right>=-J\left<\mathbf{S}_k^2\right>$. In the language of wave function of the electron pairs, we can interpret the above as a tensor of two spin wave functions with respectively electron momentum $\mathbf{k}$ and $\mathbf{Q}_{\alpha}-\mathbf{k}$. We denote here $\ket{\uparrow_{\alpha}}$ as spin up in the $\alpha$ polarization ($\alpha=x, y, z$). For example, for electron pair wave function with Cooper pair momentum $\mathbf{Q}_x$, we will have:
\begin{eqnarray}
\begin{split}
\left< S^{x}_{\mathbf{k}}\right>=&-\left< S^{x}_{\mathbf{Q}_{x}-\mathbf{k}}\right>\quad \left< S^{y}_{\mathbf{k}}\right>=\left< S^{y}_{\mathbf{Q}_{x}-\mathbf{k}}\right>\quad \left< S^{z}_{\mathbf{k}}\right>=\left< S^{z}_{\mathbf{Q}_{x}-\mathbf{k}}\right>: \\
\Rightarrow\ket{\Psi_x}&=\ket{\uparrow_x}_k\ket{\downarrow_x}_{\mathbf{Q}_x-k}+\ket{\downarrow_x}_k\ket{\uparrow_x}_{\mathbf{Q}_x-k}\\
&=\frac{1}{2}[(\ket{\uparrow_z}_k-\ket{\downarrow_z}_k)(\ket{\uparrow_z}_{\mathbf{Q}_x-k}+\ket{\downarrow_z}_{\mathbf{Q}_x-k})+(\ket{\uparrow_z}_k+\ket{\downarrow_z}_k)(\ket{\uparrow_z}_{\mathbf{Q}_x-k}-\ket{\downarrow_z}_{\mathbf{Q}_x-k})]\\
&=\ket{\uparrow_z}_{\mathbf{k}}\ket{\uparrow_z}_{\mathbf{Q}_x-\mathbf{k}}-\ket{\downarrow_z}_{\mathbf{k}}\ket{\downarrow_z}_{\mathbf{Q}_x-\mathbf{k}}
\end{split}
\end{eqnarray}

In very much the same way, we have wave functions for electron pairs with Cooper pair momentum $\mathbf{Q}_y$ and $\mathbf{Q}_z$:
\begin{eqnarray}\label{wave_function}
\begin{split}
\left< S^{y}_{\mathbf{k}}\right>=&-\left< S^{y}_{\mathbf{Q}_{y}-\mathbf{k}}\right>\quad \left< S^{z}_{\mathbf{k}}\right>=\left< S^{z}_{\mathbf{Q}_{y}-\mathbf{k}}\right>\quad \left< S^{x}_{\mathbf{k}}\right>=\left< S^{x}_{\mathbf{Q}_{y}-\mathbf{k}}\right>: \\
\Rightarrow\ket{\Psi_y}&=\ket{\uparrow_y}_k\ket{\downarrow_y}_{\mathbf{Q}_y-k}+\ket{\downarrow_y}_k\ket{\uparrow_y}_{\mathbf{Q}_y-k}\\
&=\frac{1}{2}[(\ket{\uparrow_z}_k-i\ket{\downarrow_z}_k)(\ket{\uparrow_z}_{\mathbf{Q}_y-k}+i\ket{\downarrow_z}_{\mathbf{Q}_y-k})+(\ket{\uparrow_z}_k+i\ket{\downarrow_z}_k)(\ket{\uparrow_z}_{\mathbf{Q}_y-k}-i\ket{\downarrow_z}_{\mathbf{Q}_y-k})]\\
&=\ket{\uparrow_z}_{\mathbf{k}}\ket{\uparrow_z}_{\mathbf{Q}_y-\mathbf{k}}+\ket{\downarrow_z}_{\mathbf{k}}\ket{\downarrow_z}_{\mathbf{Q}_y-\mathbf{k}}\\
\left< S^{z}_{\mathbf{k}}\right>=&-\left< S^{z}_{\mathbf{Q}_{z}-\mathbf{k}}\right>\quad \left< S^{x}_{\mathbf{k}}\right>=\left< S^{x}_{\mathbf{Q}_{z}-\mathbf{k}}\right>\quad \left< S^{y}_{\mathbf{k}}\right>=\left< S^{y}_{\mathbf{Q}_{z}-\mathbf{k}}\right>: \\
\Rightarrow\ket{\Psi_z}&=\ket{\uparrow_z}_k\ket{\downarrow_z}_{\mathbf{Q}_z-k}+\ket{\downarrow_z}_k\ket{\uparrow_z}_{\mathbf{Q}_z-k}.
\end{split}
\end{eqnarray}

If we express the above wave function (Eq. \ref{wave_function}) with electron creation operators, we will obtain the spin-triplet pairing in Eq. \ref{order_parameter_pairing}:
\begin{equation}\label{suppl_mat_order_parameter_pairing}
\begin{cases}
\hat{\Delta}_{x\mathbf{Q}_x}(\mathbf{k})=c_{\mathbf{k}\uparrow}^{\dagger}d_{\mathbf{Q}_x-\mathbf{k}\uparrow}^{\dagger} -c_{\mathbf{k}\downarrow}^{\dagger}d_{\mathbf{Q}_x-\mathbf{k}\downarrow}^{\dagger}\\
\hat{\Delta}_{y\mathbf{Q}_y}(\mathbf{k})=i(c_{\mathbf{k}\uparrow}^{\dagger}d_{\mathbf{Q}_y-\mathbf{k}\uparrow}^{\dagger}+c_{\mathbf{k}\downarrow}^{\dagger}d_{\mathbf{Q}_y-\mathbf{k}\downarrow}^{\dagger})\\
\hat{\Delta}_{z\mathbf{Q}_z}(\mathbf{k})=-(c_{\mathbf{k}\uparrow}^{\dagger}d_{\mathbf{Q}_z-\mathbf{k}\downarrow}^{\dagger}+c_{\mathbf{k}\downarrow}^{\dagger}d_{\mathbf{Q}_z-\mathbf{k}\uparrow}^{\dagger})
\end{cases}
\end{equation}

We have checked numerically that there exists uniform nearest-neighbor hopping amplitudes $\chi_{ij}^{\alpha}=\left<c_{i\sigma}^{\dagger}d_{j\sigma'}\tau_{\sigma\sigma'}^{\alpha}+h.c.\right>$ induced by the Kitaev-Heisenberg coupling on the link $r_{\alpha}$ ($\alpha=x, y, z$). This spin-dependent amplitudes renormalizes the spin-orbit coupling, however such renormalization is negligible ($J_2\chi_{ij}^{\alpha}\simeq 0.01 t'$) which justifies the mean-field decomposition in Eq.~\ref{mean_field}.

\subsection{Gor'kov-Green function}

Here, we detail the construction of the Gor'kov-Green function and specific form of the Cooper pair vertex function given in Eq.~\ref{Landau}. We can first write down the Green function for the free electron with the wave function $\Psi_{\mathbf{k}}=(c_{\mathbf{k}\uparrow}, c_{\mathbf{k}\downarrow}, d_{\mathbf{k}\uparrow}, d_{\mathbf{k}\downarrow})$:
\begin{equation}
G_0(\omega, \mathbf{k})=\frac{1}{i\omega-(\mathcal{H}_{0}(\mathbf{k})-\mu)}=\frac{P_I(\mathbf{k})}{i\omega-(\epsilon_I(\mathbf{k})-\mu)}
\end{equation}
in which $\mathcal{H}_0(\mathbf{k})$ is given in Eq.~\ref{band} and $\epsilon_I(\mathbf{k})$, $P_I(\mathbf{k})$ are respectively the energy dispersion and band projectors for the I-th band. The Gor'kov-Green function for the superconductor is obtained from the free electron Green function and the non-diagonal terms denote the pairing of the electrons.

Taking into account Eq.~\ref{mean_field}, we can write down Gor'kov-Green function, $G_{\alpha}(\omega, \mathbf{k}, \mathbf{Q})$ with the Nambu spinor $\Phi_{\mathbf{k}\mathbf{Q}}=(\Psi_{\mathbf{k}}, \Psi_{\mathbf{Q}-\mathbf{k}}^{\dagger})$, which couples free electron wave function with momentum $\mathbf{k}$ and $\mathbf{Q}-\mathbf{k}$:
\begin{equation}\label{free}
G_{\alpha}^{-1}(\omega, \mathbf{k}, \mathbf{Q})=\left(\begin{array}{cc} G_0^{-1}(\omega, \mathbf{k}) & -H_{\alpha\mathbf{Q}}(\mathbf{k}) \\ -H_{\alpha\mathbf{Q}}(\mathbf{k})^H & -G_0^{-1}(\omega, -\mathbf{k}+\mathbf{Q})\end{array}\right) 
\end{equation}
and $H_{\alpha\mathbf{Q}}(\mathbf{k})$ and $H_{0\mathbf{Q}}(\mathbf{k})$ are matrices describing the superconductivity pairing in the basis of $(c_{\mathbf{k}\uparrow}, c_{\mathbf{k}\downarrow}, d_{\mathbf{k}\uparrow}, d_{\mathbf{k}\downarrow})$ and $\tau^{\alpha}_{\sigma\sigma'}$ are the Pauli matrices in the spin subspace.
\begin{eqnarray}
\begin{split}
&H_{\alpha\mathbf{Q}}(\mathbf{k})=(J_2-J_1)\Delta_{\alpha\mathbf{Q}}\left(\begin{array}{cc} 0 & -ig_{\alpha}(-\mathbf{k}+\mathbf{Q})\tau^{\alpha}\tau^y\\
ig_{\alpha}(\mathbf{k})\tau^{\alpha}\tau^y & 0 \end{array} \right) \\
& H_{0\mathbf{Q}}(\mathbf{k})=(J_1-J_2)\Delta_{0\mathbf{Q}}\left(\begin{array}{cc} 0 & -i\tau^y g(-\mathbf{k}+\mathbf{Q}) \\ i\tau^y g(\mathbf{k}) & 0 \end{array} \right)
\end{split}
\end{eqnarray}

With Eq.~\ref{Landau}, one can write down the lowest (second) order correction to the free energy with the vertex function $\Gamma_{\alpha\beta}^{-1}(\mathbf{Q}, T)$ which denotes the electron-electron or the hole-hole interaction in the Cooper pair channel:
\begin{eqnarray}
\begin{split}
F_{BCS}(T)=&\sum_{\mathbf{Q}}F_{BCS}(\mathbf{Q}, T)\\
F_{BCS}(\mathbf{Q}, T)=&-\sum_{\omega, \mathbf{k}, \alpha, \beta}k_BT\Tr[G_0(\omega, \mathbf{k})H_{\alpha\mathbf{Q}}(\mathbf{k})G_0(-\omega, -\mathbf{k}+\mathbf{Q})H_{\beta\mathbf{Q}}^H(\mathbf{k})]+\sum_{\alpha=0, x, y, z}\frac{3}{4}(J_1+J_2)N_s|\Delta_{\alpha\mathbf{Q}}|^2\\
=&\lim_{\eta\rightarrow 0}\sum_{\alpha, \beta}\sum_{\omega, \mathbf{k}, I, J}\frac{1-n_f({\epsilon_I(\mathbf{k})-\mu)-n_f(\epsilon_J(\mathbf{-k+Q})-\mu )}}{\epsilon_I(\mathbf{k})+\epsilon_J(\mathbf{-k+Q})-2\mu+i\eta}\Tr[P_I(\mathbf{k})H_{\alpha\mathbf{Q}}(\mathbf{k})P_J( -\mathbf{k}+\mathbf{Q})H_{\beta\mathbf{Q}}^H(\mathbf{k})]\\
&+\frac{3}{4}(J_1+J_2)N_s\sum_{\alpha=0, x, y, z}|\Delta_{\alpha\mathbf{Q}}|^2\\
=& -\sum_{\alpha, \beta=0, x, y, z}N_s\Gamma_{\alpha\beta}^{-1}(\mathbf{Q}, T)\Delta_{\alpha\mathbf{Q}}\Delta_{\beta\mathbf{Q}}^*
\end{split}
\end{eqnarray}
in which $I, J$ and $\epsilon_I(\mathbf{k}), \epsilon_J(-\mathbf{k}+\mathbf{Q})$ are the indices and energies for the bands, and $P_I(\mathbf{k}), P_J(-\mathbf{k}+\mathbf{Q})$ the corresponding band projectors and $n_f(\epsilon_I(\mathbf{k})-\mu)$ is the Fermi-Dirac distribution. We have checked that the off-diagonal part of the inverse of the vertex function $\Gamma_{\alpha\beta}^{-1}(\mathbf{Q}, T)$ ($\alpha\neq\beta$) is negligible because of frustration in the momentum space. We denote the diagonal part as $\Gamma_{\alpha}^{-1}(\mathbf{Q}, T)\equiv\Gamma_{\alpha\alpha}^{-1}(\mathbf{Q}, T)$. $\Gamma_{\alpha}^{-1}(\mathbf{Q}, T)$ is actually connected to the static spin-triplet Cooper pair susceptibility $\chi_{\alpha}(\mathbf{Q}, T)$ of the non-interacting system:
\begin{equation}
\Gamma_{\alpha}^{-1}(\mathbf{Q}, T)=\chi_{\alpha}(\mathbf{Q}, T)-\frac{3(J_1+J_2)}{4}
\end{equation}

The critical temperature $T_c$ is determined by the condition $\Gamma_{\alpha}^{-1}(\mathbf{Q}, T)=0$. When $\Gamma_{\alpha}^{-1}(\mathbf{Q}, T)>0$ the superconductivity pairing $\Delta_{\alpha\mathbf{Q}}$ is stable  and the static spin-triplet Cooper pair response indicates the profile of the triplet Cooper pairs' condensate as plotted in Fig.~\ref{susc_fbz1} in the main text.

\subsection{Coherence length}

The FFLO superconductivity coherence length takes the form $\xi\simeq\sqrt{\frac{1}{\chi_{\alpha}(\mathbf{Q}, \mu, T)}\frac{\partial^2\chi_{\alpha}(\mathbf{Q}, \mu, T)}{\partial \mathbf{Q}^2}}|_{\mathbf{Q}=\mathbf{Q}_{\alpha}}$. The calculation of coherence length $\xi$ suggests that the Cooper pair pairing is short ranged around quarter-filling. In Fig.~\ref{susc_coherence}, we plot $\xi$ versus $\mu/t'$ (a) and versus $J_2/t'$ (b). This suggests that around quarter-filling and $U=6  (J_2/t'\simeq 2/3 )$  the electron pairing may involves longer range pairing other than the nearest-neighbor pairing plotted in Fig.~\ref{lattice} in the main text. 
\begin{figure}[htb]
\begin{center}
\includegraphics[width=0.25\linewidth]{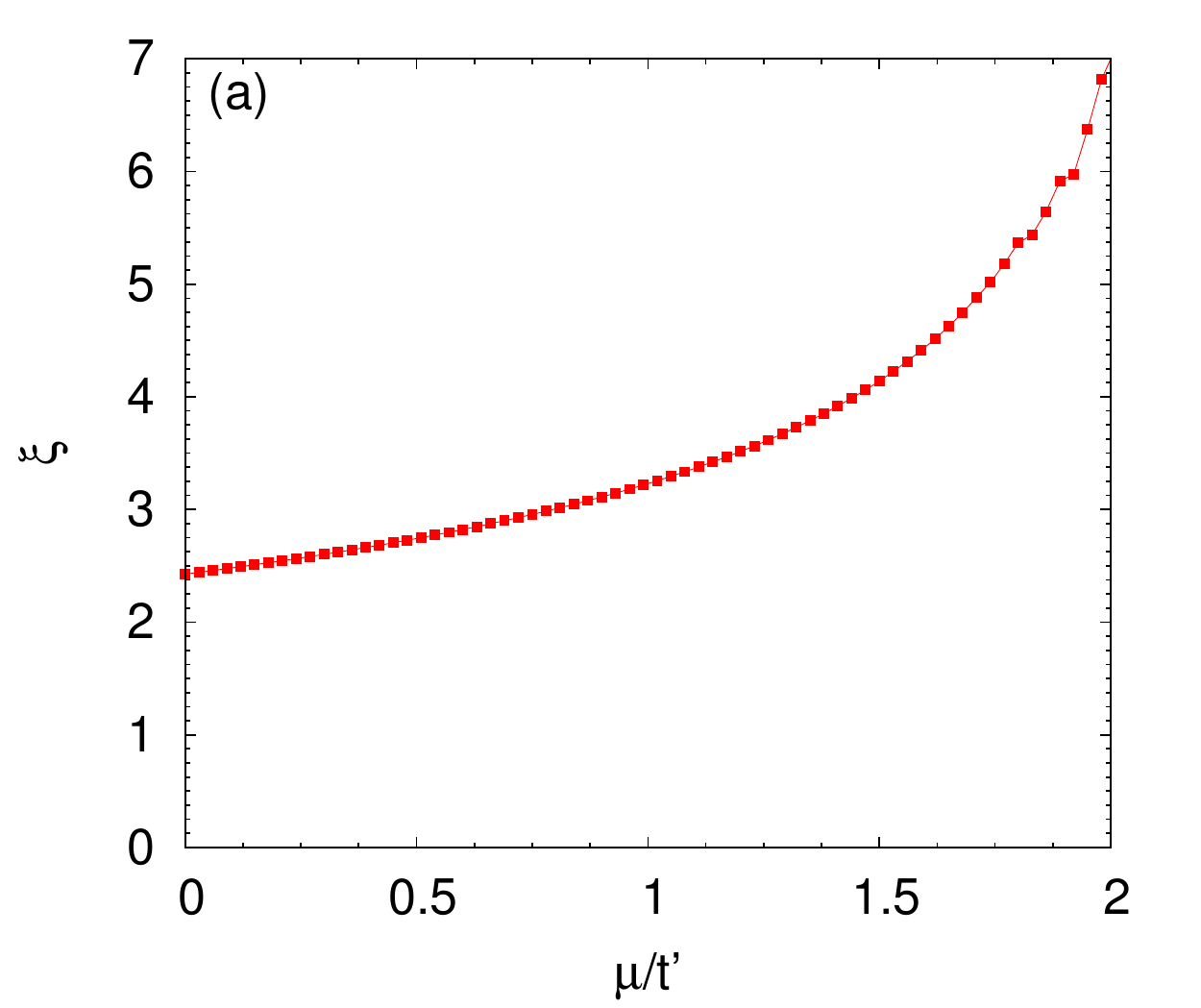}
\includegraphics[width=0.25\linewidth]{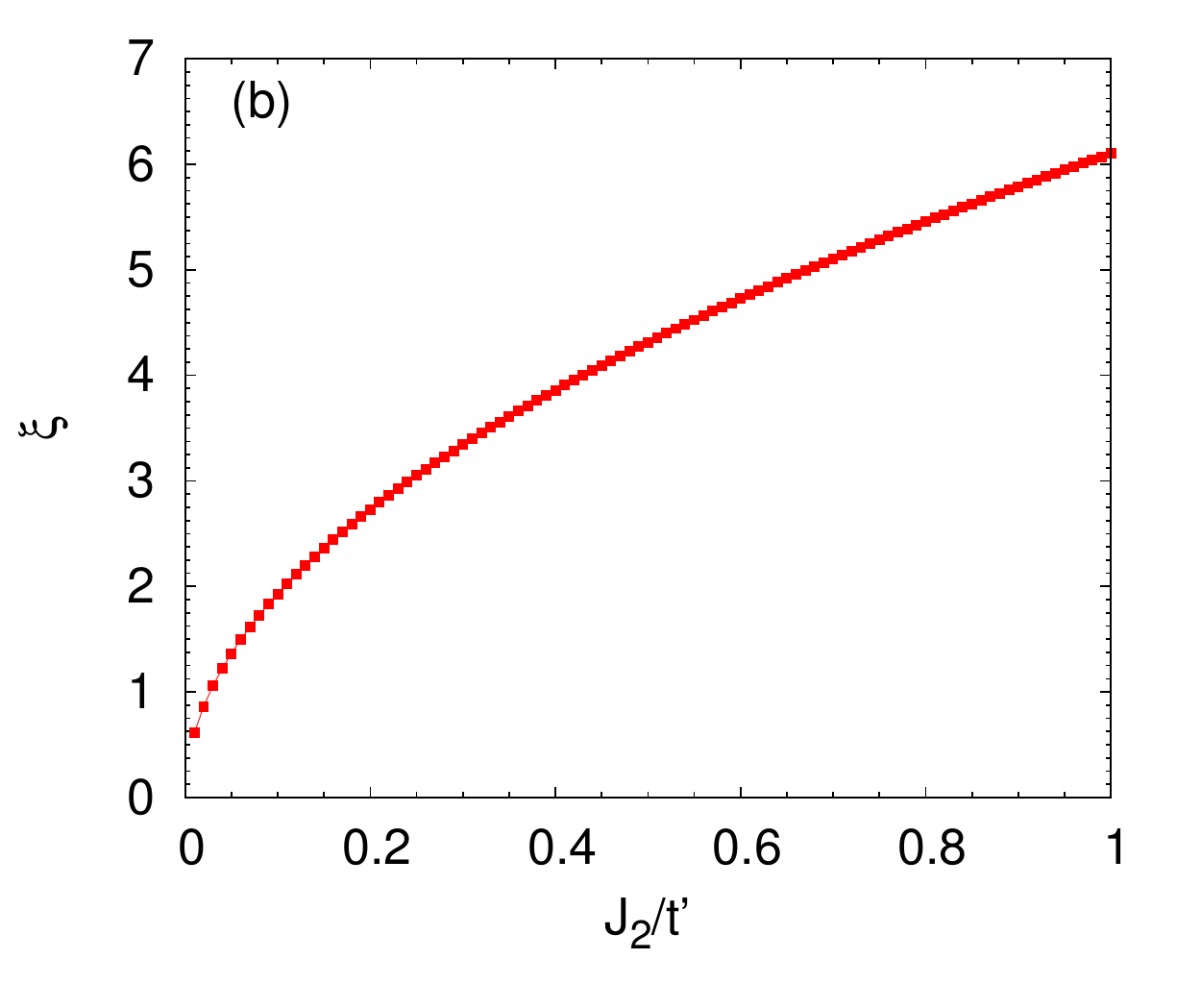}
\caption{ In the limit $t=J_1=0$: (a) The coherence length of the FFLO superconductivity $\xi$ (lattice spacing taken to be $1$) as a function of the chemical potential $\mu/t'$ at $k_BT=0.001t'$ and (b) $\xi$ as a function of the coupling $J_2/t'$ at quarter-filling $\mu=\sqrt{3}t'$, $\delta=0.25$ at $k_BT=0.001t'$.}\label{susc_coherence}
\end{center}
\end{figure}

\subsection{Bloch theorem analysis for FFLO ground state wave function}
Here, we analyze the momentum sectors of the low lying states  when the system in Eq.~\ref{Hamiltonian} is doped with a single Cooper pair. The primitive wave vectors in the direct space are denoted respectively as $\mathbf{a}_i$ (i=1, 2) with $\mathbf{a}_1=\mathbf{R}_z$, $\mathbf{a}_2=-\mathbf{R}_y$.  We numerically diagonalized the doped system and plot the energy levels in the different momentum sectors. On the discretized system of $N_x\times N_y$ plaquettes on torus, we will have thereafter $N_x$ momentum sectors in the basis of $\mathbf{k}_1=\frac{1}{N_x}(0, \frac{4\pi}{3})$ and $N_y$ momentum sectors in the basis of $\mathbf{k}_2=\frac{1}{N_y}(\frac{2\pi}{\sqrt{3}}, -\frac{2\pi}{3})$, which leads to a total of $N_x\times N_y$ sectors. To represent the 2D numerical spectra in 1D, we plot the energy levels as a function of $k_x+N_xk_y$ in which $N_x$ is the number of plaquettes in the x direction. We can apply the Bloch theorem to analyze the footprints of the FFLO superconductivity. If we denote the three degenerate ground states with one Cooper pair as $\ket{\Psi_x}$, $\ket{\Psi_y}$ and $\ket{\Psi_z}$ for the spin-triplet x, y and z, we have: 
\begin{equation}
\begin{cases}
T_1\ket{\Psi_x}=-\ket{\Psi_x} \quad T_2\ket{\Psi_x}=-\ket{\Psi_x} \quad T_1T_2\ket{\Psi_x}=\ket{\Psi_x}  \\
T_1\ket{\Psi_y}=-\ket{\Psi_y} \quad T_2\ket{\Psi_y}=\ket{\Psi_y} \quad\quad T_1T_2\ket{\Psi_y}=-\ket{\Psi_y}  \\
T_1\ket{\Psi_z}=\ket{\Psi_z} \quad\quad T_2\ket{\Psi_z}=-\ket{\Psi_z} \quad T_1T_2\ket{\Psi_z}=-\ket{\Psi_z} 
\end{cases}
\end{equation}
We have: $T_i\ket{\Psi}=e^{i\mathbf{q}\cdot\mathbf{a}_i}\ket{\Psi_i}$ in which $\mathbf{q}$ is the momentum corresponding to the momentum sector of the ground state, we can therefore identify the three ground state Ans\"{a}tze $\Psi_x$, $\Psi_y$ and $\Psi_z$: In the system with $N_x\times N_y$ plaquettes, $N_x$ and $N_y$ need to be even numbers so that the FFLO superconductivity is not frustrated. The three momentum sectors for the ground state wave function will be in $\mathbf{q}_x=(N_x/2, N_y/2)$, $\mathbf{q}_y=(N_x/2, 0)$ and $\mathbf{q}_z=(0, N_y/2)$. For a system of odd number of Cooper pairs, the total momentum equals the FFLO Cooper pair momenta are the same as the scenario with only one Cooper pair since $(2N+1)\mathbf{q}_{\alpha}\equiv \mathbf{q}_{\alpha}\mod(N_x, N_y)$ and for a system of even number of Cooper pairs, the total momentum is zero $\mathbf{q}_0\equiv 2\mathbf{q}_{\alpha}=(0, 0)\mod(N_x, N_y)$. This different behavior for ground state momentum sector between odd and even number of Cooper pairs is shown in Figs.~\ref{dop} in the main text, which distinguishes the FFLO superconductivity from other spatially modulated orders like spin or charge density wave.

\subsection{Stability of the FFLO phase when $0<t, J_1\ll t', J_2$}

We can also compute the pairing order parameter by numerical computation of the expectation value of different order parameters in the system of $2\times2$ plaquettes on torus as indicated in Fig.~\ref{lattice}:
\begin{equation}\label{order_def}
\Delta^{\alpha}_{ij}=\bra{\Psi_{N}}\hat{\Delta}^{\alpha}_{ij}\ket{\Psi_{N+2}}, \chi^{\alpha}_{ij}=\bra{\Psi_{N}}\hat{\chi}^{\alpha}_{ij}\ket{\Psi_{N}}\quad \alpha=0, x, y, z,
\end{equation}
in which the subscripts N and N+2 are the numbers of electrons for the wave functions. $\hat{\Delta}_{\alpha}$ is the triplet pairing operator given in Eq.~\ref{order_parameter_pairing} and $\hat{\chi}_{ij}^{\alpha}=c_{i\sigma}^{\dagger}d_{j\sigma'}\tau_{\sigma\sigma'}^{\alpha}+h.c.$. Using this procedure, we have calculated the superconductor pairing order parameter on different links for the $2\times2$ plaquettes on torus. The ground state for the half-filling system  ($\delta=0$) is unique while the ground state for the two electron doped system is three times degenerate. Using Eq.~\ref{order_def}, we obtain three set of order parameters and for the numerical leading order (other order parameters are at least ten times smaller), we have respectively:
\begin{equation}\label{expectation_value_order_parameter}
\begin{cases}
& 
\Delta_{16}^{x}=\Delta_{34}^{x}=-\Delta_{52}^{x}=-\Delta_{70}^{x}\\
&
\Delta_{03}^{y}=\Delta_{65}^{y}=-\Delta_{21}^{y}=-\Delta_{47}^{y}\\
&\Delta_{01}^{z}=\Delta_{45}^{z}=-\Delta_{23}^{z}=-\Delta_{67}^{z}
\end{cases}
\end{equation}
in which the numbering of the sites is represented in Fig.~\ref{lattice}. It is worth noting that the spin-triplet pairing operators  are antisymmetric: $\hat{\Delta}^{\alpha}_{ij}=-\hat{\Delta}^{\alpha}_{ji}$. The expectation values of the superconductor pairing order parameter are shown in Fig.~\ref{order}, in which we parametrize as $t'=1-t, J_1=\frac{4t^2}{U}, J_2=\frac{4t'^2}{U}, U=6.$ The numerical results confirm the emergence of triplet superconductivity with $\pi$ phase modulation for the pairing order parameter in the direct space when $t<t'$. We have checked that the charge or spin density order is negligible.
\begin{figure}[h]
\begin{center}
\includegraphics[width=0.99\linewidth]{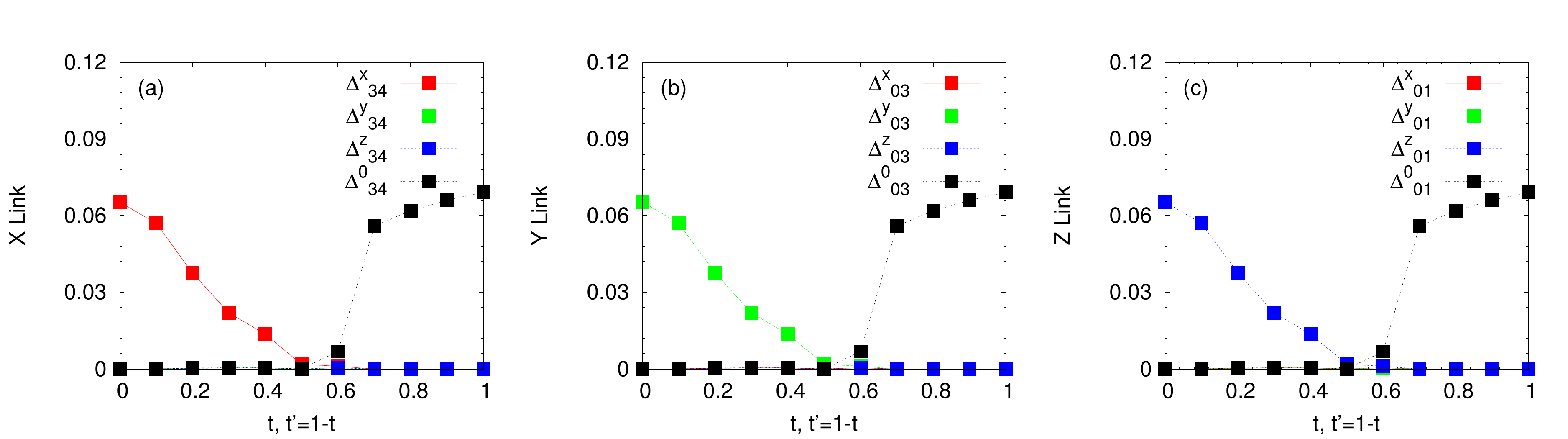}
\caption{ Order parameter calculated by equation~\ref{order_def} for the 4 plaquette system with periodic conditions: square modulus of the order parameter as a function of $t$. $t, t'=1-t, J_1=\frac{4t^2}{U}, J_2=\frac{4t'^2}{U}, U=6.$ on different links. The triplet pairing is dominant when $t<1/2$ and singlet pairing is dominant when $t>1/2$ using the ED approach. Results from the Ginzburg-Landau analysis in the thermodynamic limit suggest that the FFLO phase fade away around $t=0.3, t'=0.7$.}\label{order}
\end{center}
\end{figure}

\begin{figure}[h]
\begin{center}
\includegraphics[width=0.35\linewidth]{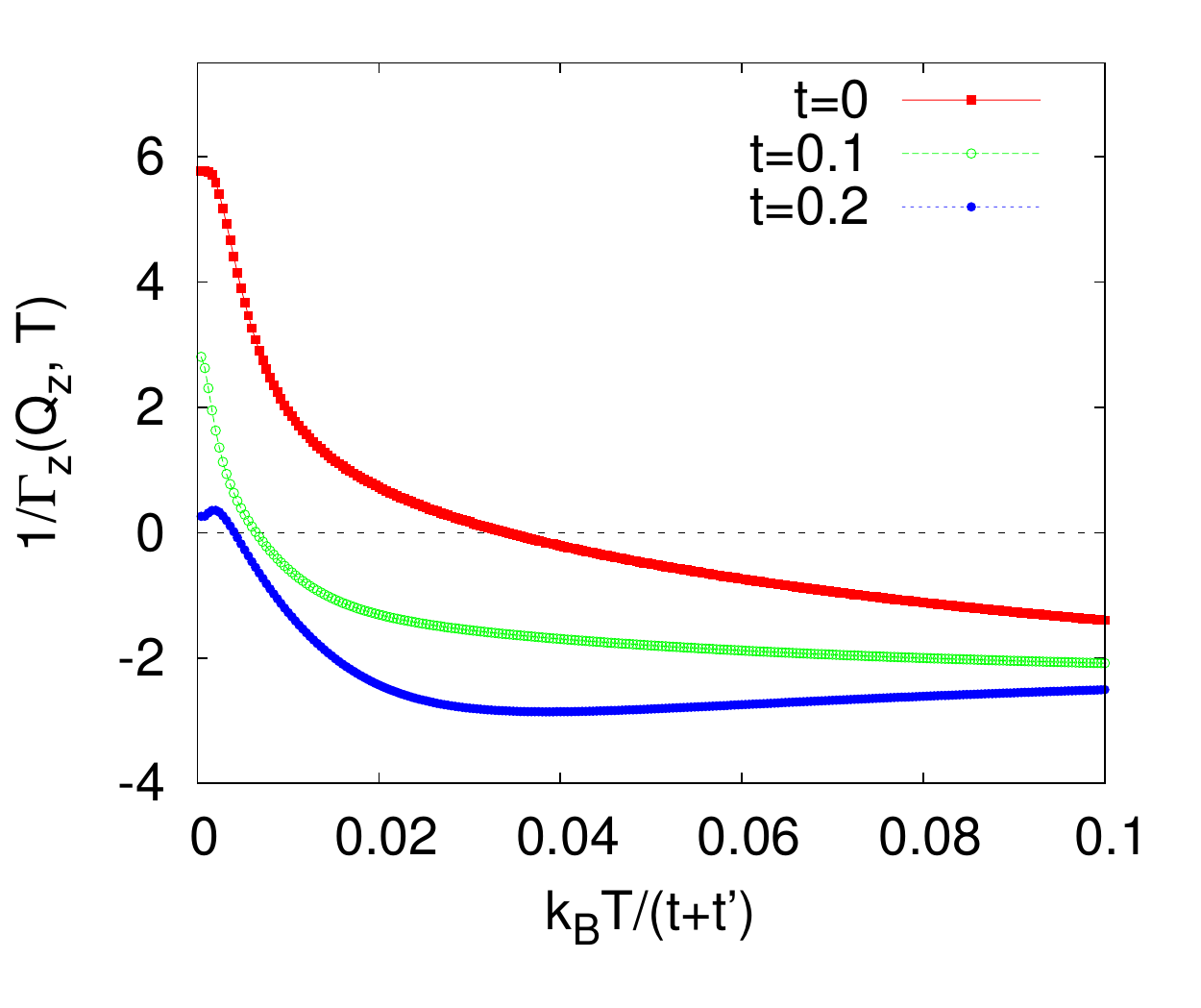}
\caption{ The peak of $1/\Gamma_{\alpha}(\mathbf{Q}_{\alpha}, T)$ as a function of temperature at $\delta=0.3$ with different values of $t$ and $t'=1-t$.}\label{susc_t}
\end{center}
\end{figure}

We have also studied the inverse of the Cooper pair vertex function for the spin-triplet pairing $\Gamma^{-1}_{\alpha}(\mathbf{Q}_{\alpha}, T)$ ($\alpha=x, y, z$) in the limit $t, J_1\rightarrow 0$. The $t$ hopping term in Eq.~\ref{Hamiltonian} opens a gap for the band structure and the linear dispersion relation becomes quadratic, which makes the susceptibility finite in the limit of zero temperature and the susceptibility becomes lower when one turns on gradually the $t$ hopping term as shown in Fig.~\ref{susc_t}. The FFLO phase appears as long as the energy related to the critical temperature is bigger than the gap of the free electron system around quarter-filling opened by the $t$ term i.e. $k_BT_c(\delta)>t$. When $t, J_1$ are large enough, the spin-singlet state will become gradually energetically favorable, which triggers the entanglement of the four Cooper pairs expressed in Eq.~\ref{order_parameter_pairing} with a term proportional to $\Delta_{0\mathbf{Q}_0}\Delta_{\alpha\mathbf{Q}_{\alpha}}\bar{\Delta}_{\beta-\mathbf{Q}_{\beta}}\bar{\Delta}_{\gamma-\mathbf{Q}_{\gamma}}$ $(\alpha\neq\beta\neq\gamma)$ in the Landau expansion.

\end{widetext}


\begin{thebibliography}{99}
\bibitem{Center symmetry}
The superconducting instability is most significant around the symmetry centers  of the Fermi surface with momentum $
\mathbf{q}$ since, for any $\mathbf{k}$, electron pairs with respectively momentum $\mathbf{k}$ and $2\mathbf{q}-\mathbf{k}$ are both located on the Fermi surface therefore the population of electrons forming Cooper pairs is the largest.
\end{thebibliography}
\end{document}